\documentclass[11pt]{article}
\usepackage[a4paper, margin=2.5cm]{geometry}

\usepackage{graphicx}
\usepackage{siunitx}
\usepackage{url}
\usepackage{hyperref}
\usepackage{amsmath}
\usepackage{amssymb}
\usepackage{amsfonts}
\usepackage{bm}
\usepackage{xspace}
\usepackage{multirow}
\usepackage{multicol}
\usepackage{xcolor}

\usepackage{authblk}

\newif\ifdoubleblind
\doubleblindfalse

\newcommand{\mm}{\si{\milli\metre}}
\newcommand{\micron}{\si{\micro\metre}}

\newcommand{\xset}[1]{\ensuremath{\mathcal{X}_{#1}}\xspace}
\newcommand{\xtrain}{\xset{tr}}
\newcommand{\xtest}{\xset{te}}
\newcommand{\xun}{\xset{un}}

\newcommand{\graph}{\ensuremath{\mathcal{G}}}
\newcommand{\nodes}{\ensuremath{\mathcal{V}}}
\newcommand{\nodestrain}{\ensuremath{\nodes{}_{tr}}}
\newcommand{\nodestest}{\ensuremath{\nodes{}_{te}}}
\newcommand{\nodesun}{\ensuremath{\nodes{}_{un}}}
\newcommand{\edges}{\ensuremath{\mathcal{E}}}
\newcommand{\node}{\ensuremath{u}}
\newcommand{\pnode}{\ensuremath{p_\node{}}}

\newcommand{\netbase}{\texttt{base}}
\newcommand{\netbasew}{\texttt{base[w]}}

\newcommand{\netres}{\texttt{res}}
\newcommand{\netresw}{\texttt{res[w]}}
\newcommand{\netresl}{\texttt{res[l]}}
\newcommand{\netreswl}{\texttt{res[w,l]}}
\newcommand{\netmlp}{\texttt{MLP}}
\newcommand{\netsage}{\texttt{SAGE}}
\newcommand{\netsaget}{\texttt{\netsage{}[3]}}
\newcommand{\netsagetr}{\texttt{\netsage{}[3+r]}}
\newcommand{\netsagefr}{\texttt{\netsage{}[5+r]}}
\newcommand{\netgat}{\texttt{GAT}}
\newcommand{\netgatt}{\texttt{\netgat{}[3]}}
\newcommand{\netgattr}{\texttt{\netgat{}[3+r]}}
\newcommand{\netgatfr}{\texttt{\netgat{}[5+r]}}

\newcommand{\conv}[3]{\ensuremath{\texttt{CONV}(#1, #2, #3)}}
\newcommand{\maxpool}[2]{\ensuremath{\texttt{MP}(#1, #2)}}
\newcommand{\fc}[1]{\ensuremath{\texttt{FC}(#1)}}

\newcommand{\cyto}{\texttt{CY}}
\newcommand{\pmap}{\texttt{PM}}
\newcommand{\coord}{\texttt{CO}}

\newcommand{\fref}[1]{Fig.~\ref{#1}}
\newcommand{\tref}[1]{Table~\ref{#1}}

\graphicspath{{images/}}

\begin{document}
%
\title{2D histology meets 3D topology: Cytoarchitectonic brain mapping with Graph Neural Networks}
%
%
\author[1,2]{Christian Schiffer}
\author[3]{Stefan Harmeling}
\author[1,4]{Katrin Amunts}
\author[1,2]{Timo Dickscheid}

\affil[1]{Institute of Neuroscience and Medicine (INM-1), Research Centre Jülich, Germany}
\affil[2]{Helmholtz AI, Research Centre Jülich, Germany}
\affil[3]{Heinrich Heine University, Düsseldorf, Germany}
\affil[4]{C\'{e}cile \& Oscar Vogt Institute for Brain Research, University Hospital Düsseldorf, Germany}

\date{}
\maketitle              
\begin{abstract}
	Cytoarchitecture describes the spatial organization of neuronal cells in the brain, including their arrangement into layers and columns with respect to cell density, orientation, or presence of certain cell types.
	It allows to segregate the brain into cortical areas and subcortical nuclei, links structure with connectivity and function, and provides a microstructural reference for human brain atlases.
	Mapping boundaries between areas requires to scan histological sections at microscopic resolution.
	While recent high-throughput scanners allow to scan a complete human brain in the order of a year, it
	is practically impossible to delineate regions at the same pace using the established gold standard method.
	Researchers have recently addressed cytoarchitectonic mapping of cortical regions with deep neural networks, relying on image patches from individual 2D sections for classification.
	However, the 3D context, which is needed to disambiguate complex or obliquely cut brain regions, is not taken into account.
	In this work, we combine 2D histology with 3D topology by reformulating the mapping task as a node classification problem on an approximate 3D midsurface mesh through the isocortex. 
	We extract deep features from cortical patches in 2D histological sections which are descriptive of cytoarchitecture, and assign them to the corresponding nodes on the 3D mesh to construct a large attributed graph.
	By solving the brain mapping problem on this graph using graph neural networks, we obtain significantly improved classification results.
	The proposed framework lends itself nicely to integration of additional neuroanatomical priors for mapping.

\end{abstract}

\section{Introduction}%
\label{sec:introduction}

Cytoarchitectonic areas are characterized by a distinct spatial organization of neuronal cells in the brain, including their arrangement into layers and columns, as well as density, orientation, or presence of certain types of cells.
As indicators for connectivity and functional modules, they provide an important microstructural reference for human brain atlases~\cite{Amunts2020}.
Mapping the borders of cytoarchitectonic areas requires to analyze brain sections at microscopic resolution.
To capture the human brain's considerable variability, it needs to be performed in many different brain samples.
\cite{Amunts2020} created a comprehensive probabilistic cytoarchitectonic atlas based on delineations in serial histological sections stained for cell bodies in a sample of ten brains, using image analysis and statistical tools to identify architectonic borders~\cite{Schleicher1999}.
These resulting maps are aggregated in a 3D reference space at the millimeter scale.
Although today's high-throughput scanners allow to digitize complete human brains at micrometer resolution, it is practically impossible to scale this approach for doing delineations in all sections of a human brain, which may have, in dependence of the size, approx. 6000-8000 sections.
This motivates the development of automatic mapping algorithms.

Recent work formulated cytoarchitectonic mapping of cortical regions as an image segmentation~\cite{Spitzer2017,Spitzer2018} or classification~\cite{Schiffer2021} problem, which can be approached with deep convolutional networks.
These methods process each 2D section individually, responding to the lack of routine workflows for computing a precise 3D reconstruction from histology en par with high throughput imaging.
In fact, as of today only one 3D reconstruction of a whole human brain from histology is available~\cite{Amunts2013}, with a spatial resolution of 20 micrometer isotropic.
The 3D topology of a brain is thus not used by these models, although it is key to disambiguate complex or obliquely cut brain regions.
Other authors suggested to incorporate inter-slice information using 3D convolutions~\cite{Milletari2016} or recurrent networks~\cite{Chen2016}, which again requires a precise 3D reconstruction.

Here we present a novel paradigm for cytoarchitectonic brain mapping which overcomes the above restrictions.
The basic idea is to reformulate the mapping task as a node classification problem on the approximate 3D midsurface mesh through the isocortex.
Since the mesh is not assumed to be precise, it can be derived from a simple linear reconstruction which is straightforward to compute from a histological image stack, and can handle a significant amount of missing sections.
Building on~\cite{Schiffer2021}, we extract deep features from 2D cortical patches at microscopic resolution using convolutional neural networks (CNNs) that were trained with a contrastive learning task to encode cytoarchitectonic characteristics.
These features are then assigned to the corresponding nodes on the reconstructed surface mesh to construct a large, attributed graph.
The brain mapping problem is solved on this graph using graph neural networks (GNNs)~\cite{Kipf2017,Hamilton2017}.

We make the following contributions:
1) We introduce a novel deep learning approach for cytoarchitecture classification in large stacks of whole brain sections which integrates high-resolution 2D texture features with global 3D topology.
2) The approach outperforms the state of the art on DL-based cytoarchitectonic mapping on a dataset of histological sections from eight postmortem human brains.
3) The approach allows flexible integration of neuroanatomical priors into the mapping problem, which further boosts classification performance.

\section{Methods}%
\label{sec:methods}

%
The proposed framework consists of three components:
1) A linear 3D reconstruction of the histological stack to compute an approximate midsurface mesh of the isocortex, which is interpreted as a graph.
2) A CNN trained with contrastive learning which extracts cortical features from the histological sections, serving as node attributes in the graph.
3) A GNN to label each graph node with a cytoarchitectonic area, exploiting both high-resolution texture and approximate 3D topology encoded in the graph.

\begin{figure}[t]
	\centering
	\includegraphics[width=0.93\linewidth]{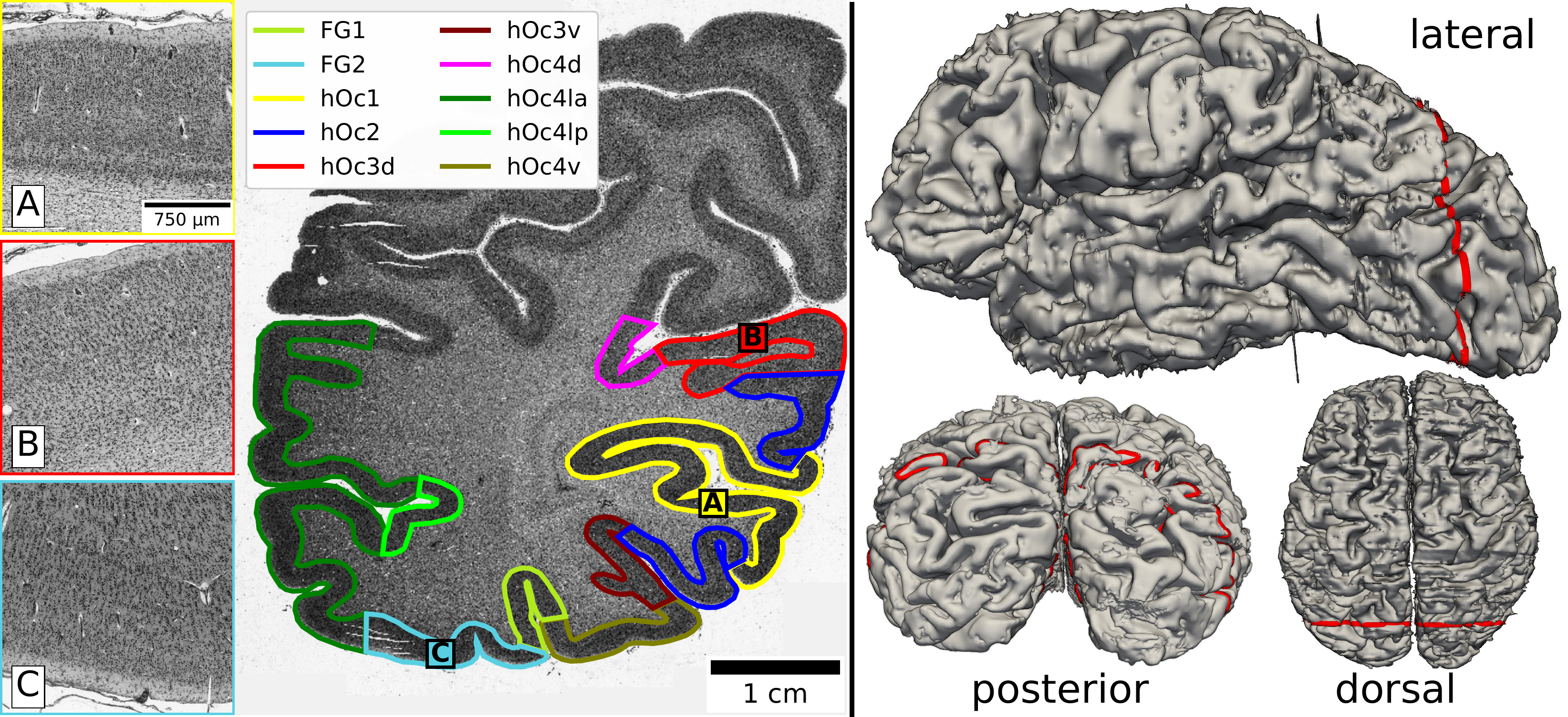}
	\caption{\textbf{Left:} Histological brain section from brain \texttt{B01} (see also \fref{fig:results}, left) with expert annotations of cytoarchitectonic areas shown in different colors.
		\textbf{Right:} Approximate midsurface through the isocortex extracted from a linear reconstruction of the histological stack.
	Points on the surface corresponding to points in the image on the left are shown in red.
}%
	\label{fig:overview}
\end{figure}

\paragraph{Construction of the approximate midsurface graph}%

\label{ssub:graph_construction_from_histological_sections}

A coarse segmentation of each section into background, white matter and cortical gray matter is performed using morphological active contours~\cite{MarquezNeila2014}.
Histological brain sections (\fref{fig:overview}, left) are 3D reconstructed by rigid alignment of adjacent sections using feature based registration as described in~\cite{Dickscheid2019}.
From the segmented linear 3D reconstruction, a 3D cortical midsurface mesh is extracted by solving Laplace's equation inside the cortical mantle~\cite{Leprince2015} using the \textit{highres-cortex} package\footnote{\url{https://github.com/neurospin/highres-cortex}} and extracting the 0.5 isosurface from the resulting volumetric Laplacian fields using marching cubes~\cite{Lewiner2003}.
In a manual postprocessing stage, hemispheres are split, and small disconnected components as well as cerebellum and brainstem are removed from the mesh.
Finally, isotropic explicit remeshing is applied using \textit{meshlab}\cite{Cignoni2008} to unify edge lengths in the mesh to $\approx 300\micron$, thereby reducing the number of triangles and ensuring that connections between vertices represent comparable distances.
Each resulting mesh (\fref{fig:overview}, right) is then interpreted as a graph $\graph{} = \left(\nodes{}, \edges{}\right)$ with nodes \nodes{} and edges \edges{}.

\paragraph{Computing node level texture features}%
\label{ssub:computing_node_level_cytoarchitectonic_features}

Each node $\node{} \in \nodes$ in the resulting graph can be uniquely identified with an image coordinate \pnode{} in a 2D histological section by inverting the rigid transformation applied for 3D reconstruction.
We apply an encoder CNN $f$ to extract a cytoarchitectonic feature embedding $\bm{h}_u$ from an image patch $\bm{x}$ centered at each coordinate \pnode{}.
The encoder $f$ is trained using the contrastive approach proposed in~\cite{Schiffer2021} to produce an embedding that maps image patches  from the same brain area to similar feature vectors, and image patches  from different brain areas to dissimilar feature vectors.
Given an image patch $\bm{x}$, $f$ produces a lower dimensional vector $\bm{h} = f(\bm{x}) \in \mathbb{R}^{D_e}$, which is passed through a projection network $g(\bm{h}) = \tilde{\bm{z}} \in \mathbb{R}^{D_p}$ and normalized as  $\bm{z}_i = \tilde{\bm{z}}_i / \left\lVert \tilde{\bm{z}}_i \right\rVert_2$.
Given a minibatch of $N$ image patches $\bm{x}_i$ with corresponding labels $y_i$ ($i=1,\ldots,N$), the contrastive loss
\begin{equation}
	\mathcal{L} = \frac{1}{N} \sum_{i=1}^N \mathcal{L}_i
	\label{eq:contrastive_loss}
\end{equation}
\begin{equation}
	\mathcal{L}_i = -\frac{1}{N_{y_i}} \sum_{j=1}^N \mathbb{I}_{i \neq j}  \mathbb{I}_{y_i = y_j} \log \frac{e^{ \langle \bm{z}_i, \bm{z}_j \rangle / \tau} }{\sum_{k=1}^N \mathbb{I}_{i \neq k} e^{ \langle \bm{z}_i, \bm{z}_k \rangle / \tau} }
\end{equation}
is optimized during training, where $\mathbb{I}$ is the indicator function, $\tau \in \mathbb{R}$ is a temperature scaling parameter and $N_{y_i}$ is the number of batch elements with the same label as $\bm{x}_i$.
The projection network $g$ is discarded after training.

\paragraph{Graph neural networks}%
\label{ssub:graph_neural_networks}

We consider two GNN architectures to integrate spatial relationships of image patches into the classification model: GraphSAGE~\cite{Hamilton2017} and Graph Attention Network (GAT)~\cite{Velickovic2018}.
Both architectures are suitable for inductive tasks, so that trained networks can be applied to previously unseen nodes or graphs.
Due to common memory constraints, training on full graphs is not possible given the level of detail of our models. We therefore adapt the neighborhood sampling scheme proposed in \cite{Hamilton2017}:
Given a node $\node \in \nodes$ and a GNN with $K$ layers, we sample its $K$-hop neighborhood $\mathcal{N}_K(\node)$ and use the resulting subgraph as input for the GNN.
This way, information from the immediate neighborhood of $\node$ can be propagated through the GNN to classify $\node$.

\paragraph{Integrating anatomical priors}%
\label{sub:integration_of_prior_knowledge}

Previous work~\cite{Spitzer2017} showed that the integration of prior anatomical knowledge - there given in the form of probabilistic cytoarchitectonic maps~\cite{Amunts2020} -  can improve classification performance.
Inspired by this finding, we adapt the registration workflow from~\cite{Amunts2020} to project probabilistic maps of 152 brain areas from the MNI Colin27 reference space onto the histological sections, and assign an additional vector $\bm{h}_u^P$ to each node $\node{}$.
Each dimension of $\bm{h}_u^P$ encodes the probability to belong to a particular region of the Julich-Brain atlas~\cite{Amunts2020}.
We further consider spatial locations in Colin27 space as another anatomical prior by assigning a 3D coordinate vector $\bm{h}_u^C$ to each node $\node{}$, again using the workflow from~\cite{Amunts2020}.



\paragraph{Implementation details}%
\label{sub:implementation_details}

Models were implemented using \emph{pytorch}~\cite{Paszke2019} and \emph{pytorch-geometric}~\cite{Fey2019}.
Training was performed on the HPC clusters JURECA~\cite{Krause2018} (NVidia K80, 12GB) and JURECA-DC (NVidia A100, 40GB) using 4 to 32 GPUs.
Code will be made publicly available\footnote{\ifdoubleblind URL anonymized \else \url{https://jugit.fz-juelich.de/c.schiffer/miccai2021_2d_histology_meets_3d_topology} \fi}.

\section{Experiments and Results}%
\label{sec:experiments}


We systematically evaluate all components of the proposed approach:
1) We compare different encoder architectures for contrastive feature learning from images patches.
2) Using the best performing encoder architecture, we compare the performance of GNNs with \netsage{} and \netgat{} architectures.
3) We investigate the additional benefit of adding neuroanatomical priors for classification.

\paragraph{Dataset}
We use an in-house dataset containing images of 1860 histological sections from 7 human postmortem brains acquired from the body donor programs of the Anatomical Institute of the University of \ifdoubleblind\texttt{********************}\else Düsseldorf (Germany)\fi, and corresponding annotations of 113 cytoarchitectonic cortical areas.
Sections have an approximate thickness of $20\micron$, were marked for neuronal cell bodies using a modified Merker stain~\cite{Merker1983}, and imaged at a resolution of $1 \micron$ with a light-microscopic
scanner (TissueScope, Huron Digital Pathology Inc.).
$80\%$ of the sections are used for training (\xtrain{}), remaining sections for testing (\xtest{}).
Transferability to unseen brains is evaluated on 325 sections from an 8th brain (\xun{}).
Graph nodes \nodes{} are split into \nodestrain{}, \nodestest{} and \nodesun{}, containing nodes with corresponding points \pnode{} in \xtrain{}, \xtest{} and \xun{}, respectively.
Performance is evaluated using macro-F1 score on nodes \nodestest{} and \nodesun{}.
The study requires no ethical approvals.

\paragraph{Performance of different feature encoders}

The encoder network $f$ is trained on 1200 image patches per class sampled from $\xtrain$, oversampling small areas for class balancing.
Data augmentation steps include random pixel intensity augmentation, rotation, mirroring, translation, blurring and sharpening with parameters detailed in~\cite{Schiffer2021}.
Training is performed for 150 epochs using LARS optimizer~\cite{You2017} with constant learning rate $0.01 * N/128$, batch size $N = 4096$ ($N = 2048$ for \netreswl{} due to memory constraints) and $\tau = 0.07$.
We consider the two network architectures listed in  \tref{tab:architecture}:
\netbase{} is the architecture presented in~\cite{Spitzer2017,Spitzer2018,Schiffer2021} which we include as baseline.
\netres{} uses pre-activation residual building blocks~\cite{He2016} and is based on ResNet18.
For both architectures, we also consider wider variants with twice as many channels per layer (\netbasew{}, \netresw{}).
For \netres{}, we further investigate larger input image patches (\netresl{}, \netreswl{}).
The following image patch sizes are used ($2\micron$/pixel): $1129^2$ pixel for \netbase{}/\netbasew{}, $1024^2$ pixel for \netres{}/\netresw{}, $2048^2$ pixel for \netresl{}/\netreswl{}.
%
Performance is evaluated on \nodestest{} and \nodesun{} by training a multi-layer perceptron (\netmlp{}) (three layers à 256 hidden units, batch normalization (BN)~\cite{Ioffe2015}, ReLU) to classify brain areas from features extracted by each considered model $f$ for all nodes in $\nodestrain{}$.
\netmlp{} is trained for 100 epochs using SGD with Nesterov momentum $0.9$, constant learning rate $0.001 * N / 4096$, batch size $N = 16384$ and cross-entropy loss.
We apply dropout with probability $0.5$ and $0.25$ to the input and hidden layers, respectively.

\begin{table}[t]
	\centering
	\caption{Network architecture of \netbase{} and \netres{} models.
		\netres{} uses pre-activation residual connections as in ResNet18~\cite{He2016}.
		\netbase{} uses no padding, \netres{} uses zero-padding
	(\conv{k}{c}{s}: $c$-channel ${k\times k}$ stride $s$ convolutional layer, \maxpool{k}{s}: ${k\times k}$ stride $s$ max-pooling, \fc{d}: $d$-dimensional fully connected layer, \texttt{GAP}: global average pooling).}
	\label{tab:architecture}
	\begin{tabular}{c|c|c}
	\hline
	layer name & \netbase{} & \netres{} \\
	\hline
	conv\_1\_x & \multicolumn{2}{c}{\conv{5}{16}{4}, \conv{3}{16}{1}, \maxpool{2}{2}} \\
	\hline
	conv\_2\_x & $\conv{3}{32}{1} \times 2$, \maxpool{2}{2} & $\conv{3}{32}{1} \times 4$ \\
	\hline
	conv\_3\_x & $\conv{3}{64}{1} \times 2$, \maxpool{2}{2} & \conv{3}{64}{2}, $\conv{3}{64}{1} \times 3$ \\
	\hline
	conv\_4\_x & $\conv{3}{64}{1} \times 2$, \maxpool{2}{2} & \conv{3}{64}{2}, $\conv{3}{64}{1} \times 3$ \\
	\hline
	conv\_5\_x & $\conv{3}{128}{1} \times 2$, \maxpool{2}{2} & \conv{3}{128}{2}, $\conv{3}{128}{1} \times 3$ \\
	\hline
	conv\_6\_x & $\conv{3}{128}{1} \times 2$ & - \\
	\hline
	projection & \multicolumn{2}{c}{\texttt{GAP}, \fc{128}, \fc{128}} \\
	\hline
\end{tabular}

\end{table}

\begin{table}[t]
	\caption{Macro-F1 scores (average across three runs) obtained on \nodestest~and \nodesun. \emph{Left}:~
		Independent patch classification based on features extracted using different encoder architectures. \emph{Center:}~Node level classification on graphs using different GNN architectures and features extracted with \netreswl{} (\netmlp{} performance for comparison). \emph{Right:}~Node level classification exploiting prior knowledge using \netsagefr{}, features from \netreswl{} and different combinations of cytoarchitectonic features (\cyto{}), probabilistic maps (\pmap{}) and canonical coordinates (\coord{}).}
	\label{tab:scores}
	\begin{tabular}{l|rr}
\hline
model & \nodestest{} & \nodesun{} \\
\hline
\netbase{} & 30.62 & 11.14 \\
\netbasew{} & 36.94 & 10.40 \\
\hline
\netres{} & 35.17 & 13.47 \\
\netresl{} & 44.67 & 17.82 \\
\netresw{} & 44.08 & 13.72 \\
\netreswl{} & \textbf{50.57} & \textbf{18.99} \\
\hline
\end{tabular}

	\begin{tabular}{l|rr}
\hline
model & \nodestest{} & \nodesun{} \\
\hline
\netmlp{} & 50.57 & 18.99 \\
\hline
\netsaget{} & 66.48 & \textbf{20.75} \\
\netsagetr{} & 68.64 & 19.84 \\
\netsagefr{} & \textbf{70.18} & 20.22 \\
\hline
\netgatt{} & 66.40 & 20.17 \\
\netgattr{} & 61.98 & 20.02 \\
\netgatfr{} & 63.46 & 20.06 \\
\hline
\end{tabular}

	\begin{tabular}{l|rr}
\hline
model & \nodestest{} & \nodesun{} \\
\hline
\cyto{} & 70.18 & 20.22 \\
\cyto{}/\pmap{} & 77.06 & 30.90 \\
\cyto{}/\coord{} & 79.28 & 33.40 \\
\cyto{}/\pmap{}/\coord{} & \textbf{79.93} & 32.67 \\
\hline
\pmap{} & 47.39 & 30.43 \\
\coord{} & 56.30 & \textbf{34.19} \\
\pmap{}/\coord{} & 56.03 & 32.30 \\
\hline
\end{tabular}

\end{table}

We found that \netres{} models perform better than \netbase{} models (\tref{tab:scores} left).
Wider networks and a larger input size improve performance on \nodestest{}.
On \nodesun{}, increased input size (\netresl{}) outperforms wider networks (\netresw{}).
Best results are obtained by combining both (\netreswl{}).

\paragraph{Graph neural networks}

We create one graph per brain hemisphere, resulting in 16 graphs overall (1.2 million nodes with node degree 6, on average).
Training nodes are sampled from $\nodestrain$, inversely proportional to their class frequency to account for class imbalance.
Node features are computed using \netreswl{}.
Node labels are assigned using annotations of cytoarchitectonic areas.
We consider two basic GNN architectures:
\netsaget{} consists of three GraphSAGE layers with mean aggregation~\cite{Hamilton2017}, while \netgatt{} consists of three GAT layers~\cite{Velickovic2018}.
Following~\cite{Li2019e,Li2020b}, we further investigate variants with pre-activation residual connections (\netsagetr{}, \netgattr{}) and five layers (\netsagefr{}, \netgatfr{}).
All layers use 256 hidden units, BN and ReLU activation.
GAT layers use 8 attention heads with 32 units each (256 units in total).
Remaining training parameters are identical to those used for \netmlp{}.
For \netsage{} models, we adapt fixed-size neighborhood sampling from~\cite{Hamilton2017} with a neighborhood size of three.
For \netgat{} models, we sample full neighborhoods, but apply dropout with probability $0.5$ to attention coefficients as proposed in~\cite{Velickovic2018}.
Both sampling methods aim to make models robust against missing nodes or edges.

All GNN architectures obtain significantly higher scores than \netmlp{} on both \nodestest{} and \nodesun{} (\tref{tab:scores}, center).
\netsagefr{} achieves overall best scores, with an increase by $19.61$ points and $1.23$ points on \nodestest{} and \nodesun{} compared to \netmlp{}, respectively.
Using the same number of layers and residual connections, \netsage{} outperforms \netgat{} in almost all cases.

\paragraph{Incorporating prior knowledge}

We examine classification performance under all possible combinations of using cytoarchitectonic features extracted from images (\cyto{}), weights from probabilistic cytoarchitectonic maps (\pmap{})~\cite{Amunts2020}, and canonical 3D locations (\coord{}) at each node.
Dropout with probability $0.5$ and additive Gaussian white noise with standard deviation $0.05$ is applied to $\bm{h}_u^P$ and $\bm{h}_u^C$, respectively.
$\bm{h}_u^P$ and $\bm{h}_u^C$ are each projected by a fully connected layer (256 units, BN, ReLU) before concatenation.

\begin{figure}[t]
	\centering
	\includegraphics[width=0.92\linewidth]{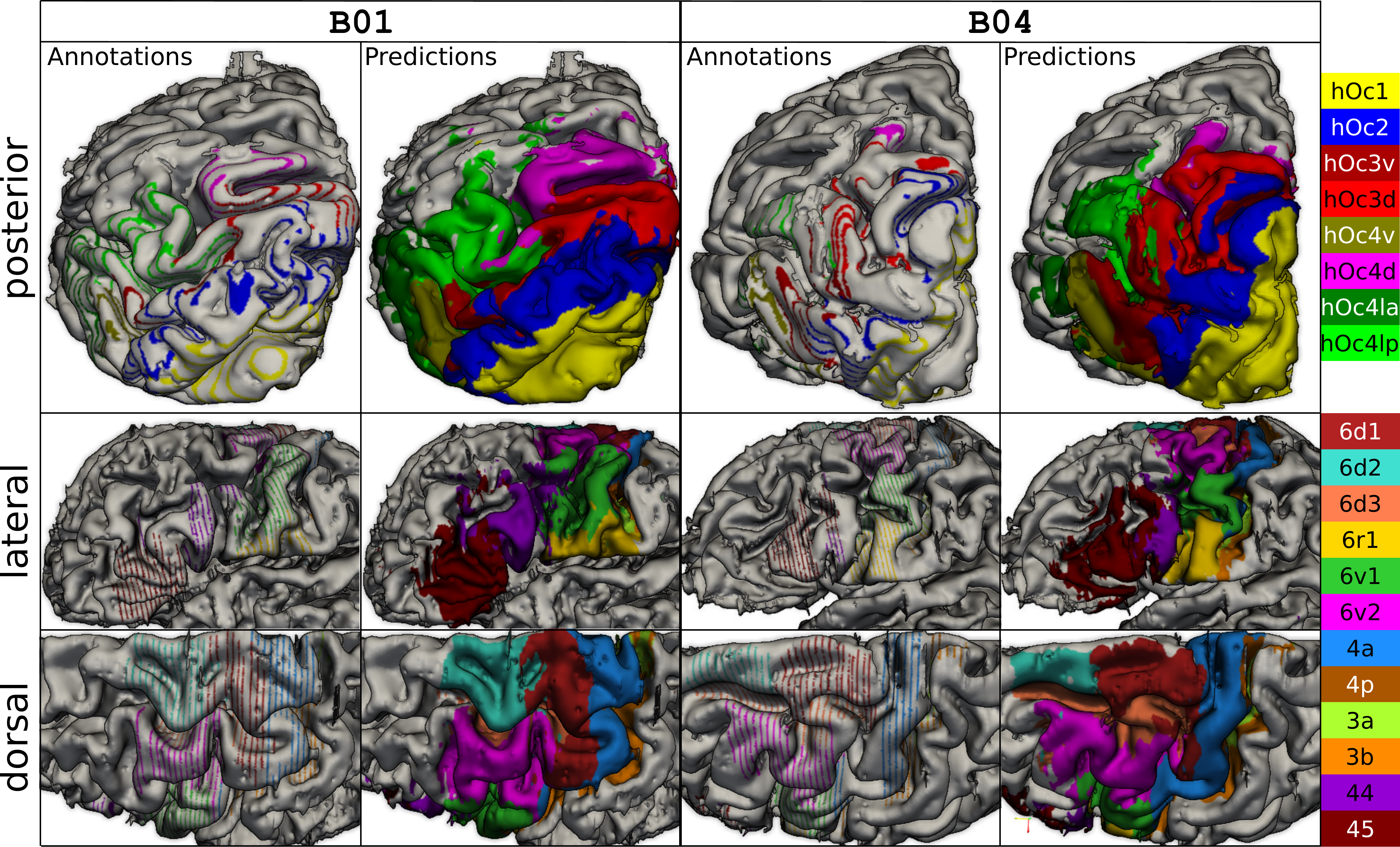}
	\caption{Visualization of available expert annotations and predictions by \netsagefr{} for the left hemisphere of two brains (\texttt{B01} and \texttt{B04}). The model was trained using features from \netreswl{} (\cyto{}) as well as anatomical priors in the form of probabilistic maps (\pmap{}) and canonical coordinates (\coord{}).
	Annotations include both training and test labels. Since they are only available for a subset of sections, they appear as stripes.}%
	\label{fig:results}
\end{figure}

Incorporating both \pmap{} and \coord{} with the texture features \cyto{} improves scores by $9.75$ points on \nodestest{} and $12.45$ points on \nodesun{} (\tref{tab:scores}, right; \fref{fig:results}).
Using them in isolation, \coord{} leads to slightly better performance than \pmap{}.
Scores on \nodestest{} drop significantly when removing texture features \cyto{}.
Best scores on \nodesun{} are obtained using only canonical coordinates, without texture features.

\section{Discussion \& conclusion}%
\label{sub:discussion_and_conclusion}

We presented a graph neural network approach for cytoarchitectonic classification of cortical image patches in microscopic scans of human brain sections, which combines 2D texture features with topological information from an approximate 3D surface reconstruction.
While \cite{Cucurull2018} employed GNNs for surface parcellation of Broca's areas 44 and 45 using MRI-based features, the present work is the first to our knowledge that integrates deep texture features from histology with 3D topology in a graph framework to label a large number of highly different brain areas across several brains.
The proposed method significantly outperforms previous methods that operate only on individual 2D sections:
The best trained model increases classification scores by $161\%$/$193\%$ on \nodestest{}/\nodesun{} wrt.~the recent baseline model \netbase{} representing the work from \cite{Schiffer2021}.
Our experiments further suggest that deeper and wider architectures for the CNN encoder $f$ benefit performance, motivating more systematic architecture search in the future.
Larger input sizes in the encoders \netresl{} and \netreswl{} showed additional positive impact by enabling models to capture the entire width of the isocortex (${2\mm\text{-}4\mm}$).
\netsage{} outperformed \netgat{}, suggesting that mean aggregation (\netsage{}) might be better suited than self-attention (\netgat{}) for our use case.
The presented framework allows straightforward integration of anatomical priors, and the results indicate that this might be a crucial strategy for optimizing  cytoarchitectonic mapping with deep networks.
Unfortunately, transferability of learned features to completely unseen brains still seems to be limited, as indicated by the performance gaps between \nodestest{} and \nodesun{} and thus confirming observations in~\cite{Schiffer2021}.
In the future, we plan to perform more rigorous architecture search for the encoders, study more anatomical priors, and investigate the reasons underlying the reduced performance on unseen brain samples.
\par
\noindent \textbf{Acknowledgements}
\ifdoubleblind
	Withheld. \\
	~\\
	~\\
	~\\
	~\\
	~\\
	~\\
	~\\
\else
	This project received funding from the European Union’s Horizon 2020 Research and Innovation Programme, grant agreement 945539 (HBP SGA3), and from the Helmholtz Association’s Initiative and Networking Fund through the Helmholtz International BigBrain Analytics and Learning Laboratory (HIBALL) under the Helmholtz International Lab grant agreement InterLabs-0015.
	Computing time was granted through JARA on the supercomputer JURECA at Jülich Supercomputing Centre (JSC).
\fi

\newpage
\bibliographystyle{splncs04}
\bibliography{library}
%




\end{document}